\documentclass[aps,prl,twocolumn,groupedaddress,nofootinbib,showpacs]{revtex4-1}

\usepackage{graphicx}
\usepackage{color}
\usepackage{amsfonts}
\usepackage{amssymb}
\usepackage{bm}

\begin{document}

\title{Polarization effects in ionic solids and melts}
\author{Mathieu Salanne$^a$}
\email{mathieu.salanne@upmc.fr}
\author{Paul A. Madden$^b$}
\affiliation{$^a$UPMC Univ Paris 06, CNRS, ESPCI, UMR 7195, PECSA, F-75005, Paris, France}
\affiliation{$^b$Department of Materials, University of Oxford, Parks Road, Oxford OX1 3PH, UK}

\begin{abstract}
Ionic solids and melts are compounds in which the interactions are dominated by electrostatic effects. However, the polarization of the ions also plays an important role in many respects as has been clarified in recent years thanks to the development of realistic polarizable interaction potentials. After detailing these models, we illustrate the importance of polarization effects on a series of examples concerning the  structural properties, such as the stabilization of particular crystal structures or the formation of highly-coordinated multivalent ions in the melts, as well as the dynamic properties such as the diffusion of ionic species. The effects on the structure of molten salts interfaces (with vacuum and electrified metal) is also described.  Although most of the results described here concern inorganic compounds (molten fluorides and chlorides, ionic oxides...), the particular case of the room-temperature ionic liquids, a special class of molten salts in which at least one species is organic, will also be briefly discussed to indicate how the ideas gained from the study of ``simple" molten salts are being transferred to these more complex systems.

\end{abstract}
\maketitle

\section{Introduction}

Molecular dynamics now is an indispensable tool for solving many physical chemistry problems. Alongside the constantly increasing efficiency of computers, the development of new modeling  strategies as well as efficient algorithms has opened the way for a new approach to the study of many systems.  Simulation may be employed as a predictive tool that can be used to complement, or even anticipate experimental measurements. Ideally, to be completely predictive, atomic-scale simulations should be based entirely on parameter-free quantum chemical calculations. This is not yet achievable, so that constructing better models for interatomic interactions remains an important goal, especially for condensed matter systems.~\cite{carter2008a} That is, to reach the time scales and system sizes which are necessary to understand complex phenomena such as transport mechanisms in the liquids or biological processes, the use of explicit interaction potentials derived from a model is necessary.~\cite{stone2008a} Computer simulations are a particularly important predictive tool for molten salts, for which experiments are often difficult or even impossible because of the extreme physical conditions and because many melts are highly corrosive.~\cite{rollet2004a,rollet2009a} The objective for model development is a {\it transferable} model, i.e. one in which the corresponding interaction potential parameters have to be determined once and for all before to be used in a variety of thermodynamic conditions (temperature, pressure, composition).

Halide melts have been the testing ground for this model development, as they were the target of pioneering experimental studies to examine the atomic scale structure by spectroscopy \cite{papatheodorou1977a,photiadis1998a,dracopoulos2001a} and diffraction \cite{ohno1978a,ohno1978b,wasse1999a}; more recently attention has shifted to oxides for which the experimental difficulties are more pronounced.  Due to the monatomic nature of the halide melts, the complexities of {\it intra}molecular interactions and molecular shape are avoided both in constructing the interaction model and in extricating the influence of intermolecular effects from experimental data. Although, as we will see, complex local coordination structures are present in the melts,~\cite{glover2004a,rollet2011a} they arise from the interplay of the atomic interactions. Most of the short-range structural properties can be attributed to the competition between the overlap-repulsion on the one hand and the Coulombic interaction on the other hand \cite{rovere1986a}. The former, which is a consequence of the Pauli principle, will determine the closest approach distances while the latter will induce strong ordering effects: Around a given ion, the first solvation shell will always be entirely constituted of oppositely charged species; this ordering automatically transfers up to several solvation shells. The third term of the intermolecular interactions, the dispersion, arises from correlated fluctuations of the electrons. It is always attractive and although it brings a smaller contribution than the other terms to the overall energy, it has important effects on the packing properties, thus influencing many thermodynamic properties such as the density or the surface tension.

In a first approximation, the total energy arising from these three terms (repulsion, Coulomb and dispersion) can be written as a sum over all the pairs of ions.\cite{tosi1964a,fumi1964a} From the simulation point of view, it is rather straightforward to implement pair-potentials, so that the first simulations of molten salts were performed relatively soon after the first molecular dynamics simulations were reported.~\cite{alder1959a,woodcock1971a,lantelme1974a,hansen1975a} Significant results were obtained from these first simulations. For example the bulk properties of several alkali halide crystals could be determined with a good precision, and a complete atomic-scale picture of the structure and transport in simple molten salt was obtained for the first time.~\cite{lantelme1982a,lantelme1984a}

In a second stage, similar interaction potentials were tested for the study of more complicated halide melts including multiply charged cations, such as Zn$^{2+}$ in ZnCl$_2$. The results were not so convincing; for example the predicted structures differed qualitatively from the experimental ones, which had been obtained from neutron or X-ray diffraction.~\cite{biggin1981a,triolo1981a,allen1991a,zeidler2010a} Although the tetrahedral structure of the Zn$^{2+}$ first-neighbour coordination shell was well reproduced, the corner-sharing arrangement between the tetrahedra was not predicted by the simulations,~\cite{woodcock1976a,gardner1985a} unless unphysically large dispersion interactions between the cations were introduced.~\cite{kumta1988a} This inconsistency was solved by introducing a fourth component into the interaction potential, which represents the polarization of the electron cloud of an ion in response to its environment.~\cite{wilson1993a,wilson1993b} This term cannot be implemented in the form of a pair-potential because of its inherent many-body character. The simulations including polarization effects on the chloride anions were able to reproduce the structure factors extracted from experiments not only for pure ZnCl$_2$ but also for its mixtures with alkali chlorides~\cite{wilson1994a}, which allowed for an understanding of the basic mechanisms for network formation in ionic fluids.

A systematic set of polarizable potentials for ionic solids based upon the Shell Model \cite{dick1958a} had been proposed by Sangster and Dixon.~\cite{sangster1976a} Yet these potentials suffered from numerical instability problems (the so-called polarization catastrophe) especially when applied to melts at high temperature and pressure conditions. Madden and Wilson proposed a different approach,~\cite{wilson1993a} which also included an important short-range polarization effect. This term was introduced as a consequence of a study of polarization in condensed phases based upon first-principles electronic structure calculations \cite{fowler1985a}- rather than as a simple empirical stabilization device to avoid the polarization catastrophe. The model which was introduced in this study is now widely used in the study of ionic materials, ranging from molten fluorides~\cite{merlet2010a,pauvert2010a} to silicates~\cite{tangney2002a,giacomazzi2011a} and silver halides.~\cite{bitrian2011a,alcaraz2011a} By relating the parameters of the interaction model to properties of the individual ions (ion size, polarizability {\it etc.}) in a chemically-inspired way the model was shown able to account systematically for the solid state structures adopted by a very wide range of materials \cite{wilson1997b}. More recently, methods have been introduced to allow most of the parameters which enter into the interaction models to be derived directly from first-principles, rather than allowing them to be determined empirically by adjusting them to reproduce the experimental properties of the simulated material \cite{salanne2011c}. These {\it ab initio}-determined properties have included polarizabilities \cite{heaton2006b}, the short-range repulsion and also the dispersion interactions \cite{rotenberg2010a}, so that the objective of finding realistic, predictive and transferable models has been accomplished, at least for halide melts.

In the first section, we will introduce the corresponding functional form; with special emphasis given to the methodological aspects associated with the polarization effects. The second part details the consequences for the physico-chemical properties which relate to structure and dynamics. The interfacial properties, which are important in many industrial applications of ionic melts, are also discussed. We focus on the systems for which most of the data are available, i.e. inorganic ionic compounds, but the particular case of the room-temperature ionic liquids, a special class of molten salts in which at least one species is organic, will also be briefly discussed to indicate how the ideas gained from the study of ``simple" molten salts are being transferred to these more complex systems. The {\it ab initio} parameterization of the potential parameters will not be considered in any detail here, it has been the subject of another recent review \cite{salanne2011c}.

\section{Introducing polarization effects in molecular dynamics simulations}

The focus of this topical review are the polarization effects themselves, so that other many-body effects that may occur in ionic liquids like the ``breathing'' of the anions will not be discussed.~\cite{madden1996a} Under this condition, the charge-charge, repulsion and dispersion contributions to the energy can be expressed through:

\begin{equation}
V^{\rm RIM}=\sum_{i<j}\left(\frac{q^iq^j}{r^{ij}}+ A^{ij}{\rm e}^{-a^{ij}r^{ij}}-f_6^{ij}(r^{ij})\frac{C_6^{ij}}{(r^{ij})^6}-f_8^{ij}(r^{ij})\frac{C_8^{ij}}{(r^{ij})^8}\right)
\label{eq:rim}
\end{equation}

\noindent where the superscript RIM stands for ``rigid ion model''. In the framework of an ionic model, which is based upon the interactions of closed-shell species, the charges $q^i$ should be the formal, valence ones (except in the special case of redox active species, which has hardly been tackled by molecular dynamics simulations~\cite{pounds2009a}). $A^{ij}$, $a^{ij}$, $C_6^{ij}$ and $C_8^{ij}$ are parameters which have to be set up for each ion pair, and $f_n^{ij}$ are Tang-Toennies dispersion damping functions,~\cite{tang1984a} describing the short-range penetration correction to the asymptotic multipole expansion of dispersion,~\cite{stone-book} which take the following form:

\begin{equation}
f_n^{ij}(r^{ij})=1-{\rm e}^{-b_n^{ij}r^{ij}}\sum_{k=0}^n\frac{(b_n^{ij}r^{ij})^k}{k!}
\label{eq:tangtoennies1}
\end{equation}
\noindent Note that in many studies this correction is not taken into account,  which is done by replacing the Tang-Toennies function by a constant value of 1.0 for any interatomic distance $r^{ij}$.

Compared to the RIM, the polarizable ion model (PIM) takes into account the effects arising from the polarization of each ion:

\begin{equation}
V^{\rm PIM}=V^{\rm RIM}+V_{\rm polarization}
\end{equation}

\noindent This polarization, which is the response of the electron cloud of the ion $i$ to the local electric field and its derivatives at the position ${\bf r}^i$, together with short-range effect to be described below, can be represented with a multipole expansion. For example, up to the second order, there will be induced dipoles and quadrupoles on each ion:
\begin{eqnarray}
\mu_\alpha^{i, as} & = & \alpha_{\alpha\beta}E_\beta({\bf r^i})+\frac{1}{3}B_{\alpha\beta\gamma\delta}E_\beta({\bf r^i})E_{\gamma\delta}({\bf r^i})+...\label{eq:multipoles} \\
\theta_{\alpha\beta}^{i, as} & = &\frac{1}{2} B_{\alpha\beta\gamma\delta}E_\gamma({\bf r^i})E_\delta({\bf r^i})+C_{\alpha\beta\gamma\delta}E_{\gamma\delta}({\bf r^i})+...
\end{eqnarray}

\noindent where ${\boldsymbol{\alpha}}$ and ${\bf C}$ are the dipole and quadrupole polarizabilities and ${\bf B}$ is the dipole-dipole-quadrupole hyperpolarizability.~\cite{buckingham1967a} $E_\alpha$ and $E_{\alpha\beta}$ are components of the electric field and field gradient, respectively.

\begin{figure}
\begin{center}
\includegraphics[width=9cm]{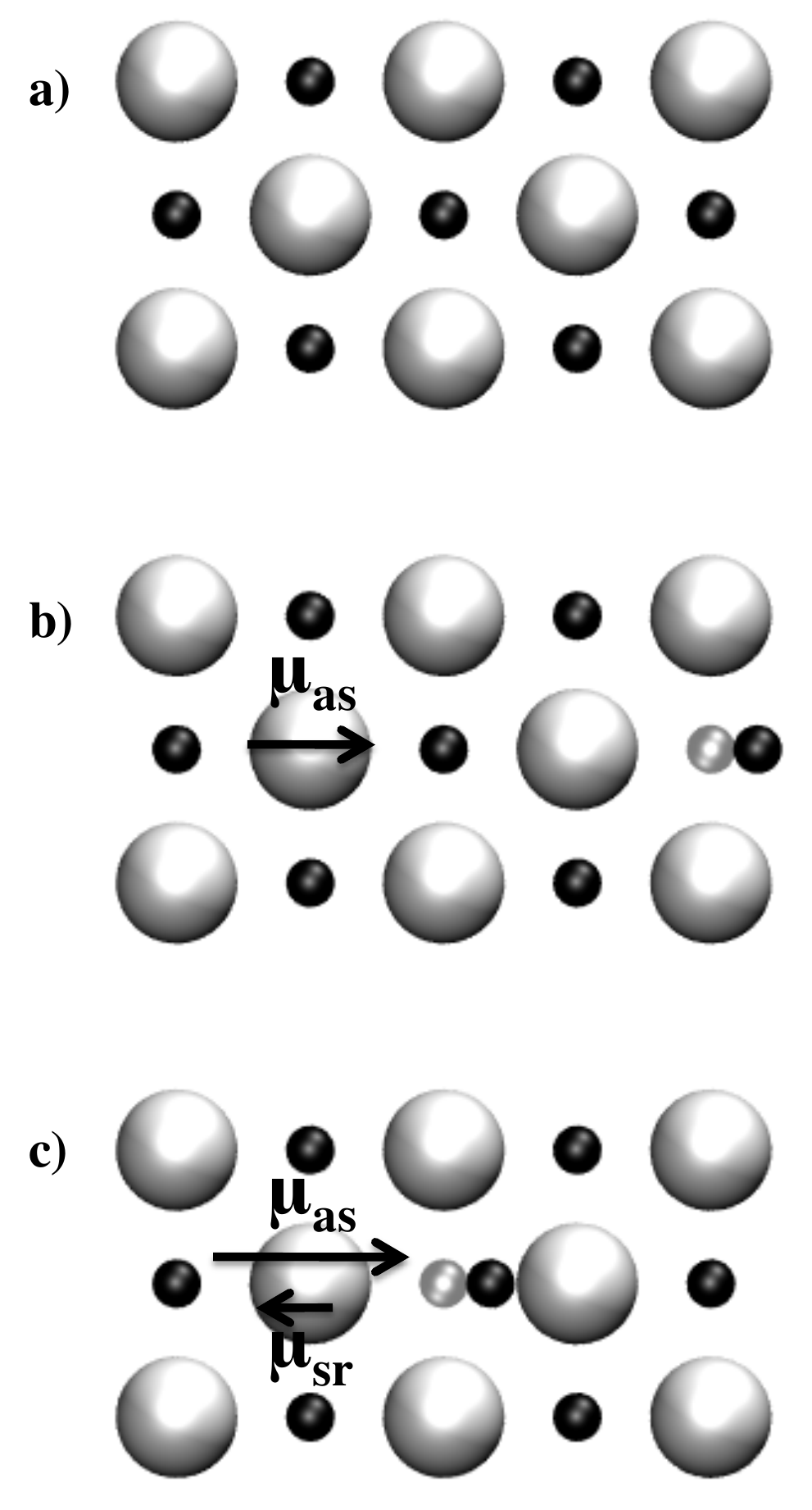}
\caption{\label{fig:dipolescristal} Origin of the ``asymptotic'' and ``short-range'' contributions to the dipole induced on an anion in a crystal. a) The crystal is perfect, the anion does not feel electric field and there is no induced dipole. b) A cation at a distance greater than next-nearest neighbour separation from the considered anion is displaced off its lattice site. The anion feels an electric field which induces an asymptotic induced dipole ${\boldsymbol{\mu}}^{i, as}$. c) A cation in the immediate vicinity of the anion is displaced off its lattice site. Whilst the electric field tends to push the electrons away from the displaced cation (${\boldsymbol{\mu}}^{i, as}$), the electrons also have more freedom to move into the space vacated, which results in a short-range contribution to the induced dipole (${\boldsymbol{\mu}}^{i, sr}$).}
\end{center}
\end{figure}

In the present manuscript, we will consider the case of dipole polarization effects only. As we shall see, in the case of many ionic species, this choice can safely be made without hindering the predictive capabilities of the model. In equation \ref{eq:multipoles}, the superscript $as$ is used because it corresponds to the $asymptotic$ contribution to the induced dipole only. Electronic structure calculations have showed that a substantial short-range effect on the induced dipole is not taken into account by this expression.~\cite{fowler1985a,jemmer1999a} The origin of this short-range contribution to the dipole is schematized on figure \ref{fig:dipolescristal}. Compared to the perfect crystal cases where no induced dipoles are created, when  a cation at a distance greater than next-nearest neighbour separation from the considered anion is displaced off its lattice site the latter feels an electric field which induces an asymptotic induced dipole ${\boldsymbol{\mu}}^{i, as}$. But when a cation in the immediate vicinity of the anion is displaced off its lattice site, whilst the electric field tends to push the electrons away from the displaced cation (${\boldsymbol{\mu}}^{i, as}$), the electrons also have more freedom to move into the space vacated (${\boldsymbol{\mu}}^{i, sr}$). The total dipole induced by displacing first neighbour cations is reduced below that expected from the asymptotic term. On the contrary, when the central, polarizable species is the cation, the asymptotic and short-range dipoles generally point towards the same direction.~\cite{domene2001a} In practical applications, the short-range effects on the induced dipoles are taken into account by again introducing Tang-Toennies damping functions~\cite{tang1984a}

\begin{equation}
g_D^{ij}(r^{ij})=1-c_D^{ij}{\rm e}^{-b_D^{ij}r^{ij}}\sum_{k=0}^n\frac{(b_D^{ij}r^{ij})^k}{k!}
\end{equation}

\noindent  where we draw the attention to the presence of an additional parameter, $c_D$, compared to  equation \ref{eq:tangtoennies1} which is used for the dispersion interaction. This parameter measures the strength of the ion response to the short-range effect and therefore depends on the identity of the ion. The efficiency of these functions in accounting for short-range effects was tested against {\it ab initio} calculations of the induced multipoles in distorted crystals.~\cite{jemmer1999a,domene2002a}

The total induced dipole on each ion then satisfies the following equation:

\begin{equation}
{\boldsymbol \mu}^i=\alpha^i \sum_{j\ne i}\left(g_D^{ji}(r^{ij}){\bf T}^{(1)}q^j-{\bf T}^{(2)}\cdot{\boldsymbol \mu}^j \right)
\label{eq:scf}
\end{equation}

\noindent where we have introduced the charge-dipole and dipole-dipole interaction tensors:
\begin{eqnarray}
{\bf T}^{(1)}&=&{\boldsymbol \nabla}\frac{1}{r^{ij}}=-\frac{1}{{r^{ij}}^3}{\bf r}^{ij}\\
{\bf T}^{(2)}&=&{\boldsymbol \nabla}\otimes{\bf T}^{(1)}=\frac{3}{{r^{ij}}^5}{\bf r}^{ij}\otimes{\bf r}^{ij}-\frac{1}{{r^{ij}}^3}{\bf I}
\end{eqnarray}
\noindent and where {\bf I} is the identity matrix. During a molecular dynamics simulation the induced dipoles have to be determined at each time step; the presence of the dipoles of all the ions other than $i$ in the right hand term of equation \ref{eq:scf} obliges us to solve the set of $N$ equations self-consistently. As soon as the dipoles are known, the polarization energy is calculated from

\begin{eqnarray}
V_{\rm polarization}&=&\sum_{i,j}\left[ \left(q^j\mu^i_{\alpha}g_D^{ji}(r^{ij})-\mu_\alpha^jq^ig_D^{ij}(r^{ij})\right)T_\alpha^{(1)}\right.\\
& & \left. -\mu_\alpha^i\mu_\beta^jT_{\alpha\beta}^{(2)}\right]  +\sum_i \left(\frac{1}{2\alpha^i}\mid {\boldsymbol \mu}^i \mid^2 \right) \nonumber
\end{eqnarray}

\noindent where the first two terms arise respectively from the charge-dipole and the dipole-dipole interaction, and the last term corresponds to the the energy cost of deforming the charge density of the ion $i$ (with polarizability $\alpha^i$) to create the induced dipole. We immediately see that solving the set of equations \ref{eq:scf} self-consistently  is formally equivalent to solving:
\begin{equation}
\left(\frac{\partial V_{\rm polarization}}{\partial {\boldsymbol \mu}^i}\right)_{\{\boldsymbol{\mu}^j\}_{j\ne i}}=0,
\label{minimization}
\end{equation}
\noindent that is, to minimize $V_{\rm polarization}$ with respect to the induced dipoles. Note that, in the simulations, the polarization effects which are intrinsically of a many-body nature, are calculated at the cost of evaluating only pairwise additive interactions (plus the cost of enforcing the minimization condition)! The minimization task can be performed by using preconditioned conjugate gradients methods, which appears to be much faster than the self-consistent approach. The dynamics is thus similar to the so-called Born-Oppenheimer {\it ab initio} molecular dynamics. As for the latter, algorithms based on the Car-Parrinello approach~\cite{car1985a,remler1990a,sprik1988a} or on predictor-corrector schemes~\cite{kolafa2004a,sala2010a} can also be used to determine the induced dipoles instead.

The polarization potential (as well as the Coulombic one) consists of long-ranged interactions, for which it is necessary to go beyond the normal minimum image truncation and to sum the intermolecular interactions over the periodic images of the simulation cell. The most popular method for doing this is the Ewald summation technique, which consists in decomposing the interaction potential into a short-range component summed in real space and a long-range component summed in Fourier space.~\cite{ewald1921a,allen-livre} Several studies were devoted to the implementation of this technique for multipole moments,~\cite{nymand2000a,aguado2003a,laino2008a} and the literature on this topic was recently reviewed by Stenhammar {\it et al.}~\cite{stenhammar2011a} These authors pointed to several discrepancies between the formulae in the published work and they provided a consistent set of expressions.

Expressions for the stress tensor and for other quantities like the heat current, which are obtained by differentiation of the expression for the total energy  with respect to external variable, are also needed in order to perform molecular dynamics simulations in different ensembles and to determine properties like the viscosity or thermal conductivity. For example, an expression for the stress tensor, required in  $NPT$ simulations or to calculate the viscosity, is obtained by differentiating the energy with respect to the shape of the simulation cell.~\cite{parrinello1980a,martyna1994a} For a system in which the interaction between the particles are described by short-range pair potentials, the stress tensor elements are given by:
\begin{equation}
\Pi_{\alpha\beta}=\sum_i m^i \dot{r}^i_\alpha \dot{r}^i_\beta-\sum_{j>i}r^{ij}_\alpha \frac{\partial V}{\partial r^{ij}_\beta}.
\end{equation}
\noindent where $m^i$ is the mass of particle $i$. Here again, special care must be taken in the calculation of the long-ranged interactions, and the Ewald summation method must be employed.~\cite{aguado2003a} In a simulation model which contains additional degrees of freedom, like the dipoles in the present polarizable potentials whose values will vary when the cell shape is changed, the question arises as to how they are to be handled in obtaining these expressions. The answer is that because the dipoles are in fact wholly determined by the instantaneous positions of all ions, because of the minimization condition (equation \ref{minimization}), any additional terms due to the derivatives of dipoles with respect to cell-shape are to be ignored.  The resulting expressions will still involve the dipole values and they must always be evaluated with the dipoles obeying the minimization condition. Once again, the expressions for the stress tensor {\it etc.} are then evaluated from only pairwise additive expressions, despite the implicit many-body character of the interactions giving rise to them.

\section{Illustration of the polarization effects}

\subsection{Polarization effects on the structure}

The first ionic compounds which were systematically studied by molecular dynamics (and Monte-Carlo) simulations were the alkali halides. For such systems, the simple RIM provides a rationalization of the crystal structures depending on the two ionic radii. In all cases the unlike ion Coulombic interactions are maximized, and the maximal number of anions (cations) around a given cation (anion) is limited by their relative size  only. Several typical structures can therefore be formed depending on their relative packing. For successively lower cation/anion radius ratio, these are the eight-coordinate caesium chloride (B2), six-coordinate rocksalt (B1) and four-coordinated blende (B3) or wurtzite (B4) structures. The mechanisms involved in the successive transformations when passing from one phase to another have also been investigated.~\cite{wilson2002a}

\begin{figure}
\begin{center}
\includegraphics[width=9cm]{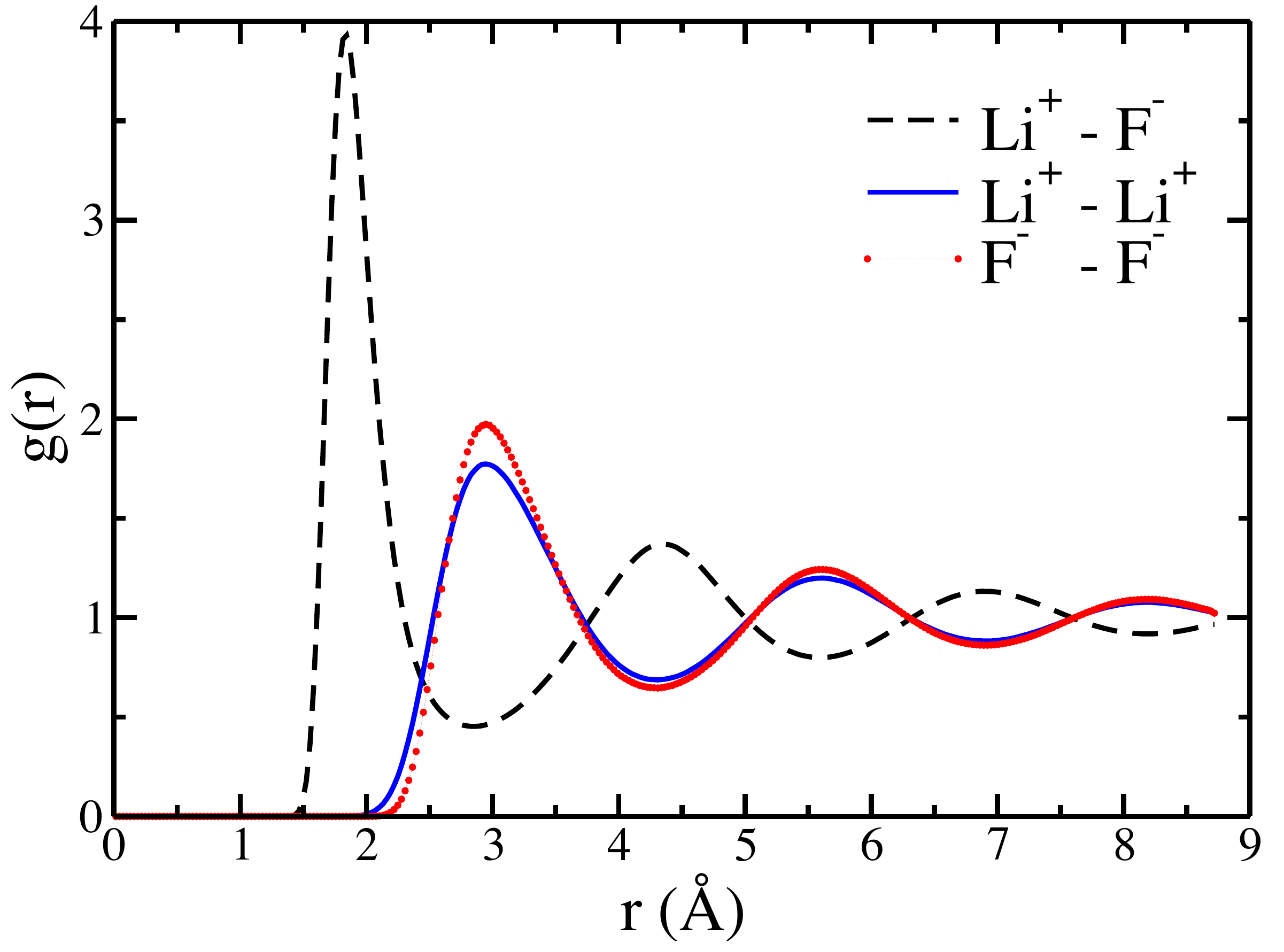}
\caption{\label{fig:rdf} Partial radial distribution functions in pure LiF at 1123~K.}
\end{center}
\end{figure}

Upon melting, alkali halides keep a structure based on a competition between the overlap-repulsion and Coulombic interactions \cite{rovere1986a}. A typical set of radial distribution functions (RDFs), obtained for pure LiF at 1200~K, is provided on figure \ref{fig:rdf}. The main feature to be noted is the position of the first maximum of the like-like RDFs, which occur at the same position coincident with the first minimum of the Li$^+$-F$^-$ one; the same behavior is also observed for all the following extrema. For all these systems, no structural change was observed when including the polarization effects.~\cite{morgan2004a}

As soon as ions with multiple charges are involved, the situation changes, there are multiple examples of multivalent metal halide species for which polarization effects have shown to be important. Simulations based on the RIM model were unable to predict the structure of many MX$_2$ compounds (where M$^{2+}$~=~Mg$^{2+}$, Zn$^{2+}$, Mn$^{2+}$, Ca$^{2+}$, Sr$^{2+}$ or Ba$^{2+}$ and X$^-$ is an halide anion). This is particularly true for compounds with a low cation/anion radius ratio, which tend to crystallize in layered CdI$_2$ or CdCl$_2$ structures despite the fact that such structures involve shorter separations between the highly charged cations than could be achieved by other ways of accommodating the cations in the anion lattice.~\cite{madden1996a} The role of polarization effects in the formation of these particular structures, which persist in the molten phase, has now been well identified. In fact all these structures depart from the picture provided by the RIM because they involve the formation of bent M-X-M angles, whereas the coulombic repulsion between the highly charged cations tends to push them as far apart as possible. The polarization is the driving force for the occurrence of this non-trivial bond angle: When the halide anion is displaced off the line of centres of the cations a dipole is induced, which serves to screen the cation-cation repulsion and lowers the total energy. Even for oxide materials, for which many-body effects other than polarization such as the aspherical breathing of the oxide anion O$^{2-}$ have to be taken into account for an accurate and transferable description~\cite{madden1996a,aguado2004a,aguado2005a,jahn2007b}, the polarization effects play a dominant role in determining these local ionic arrangements.

\begin{figure}
\begin{center}
\includegraphics[width=9cm]{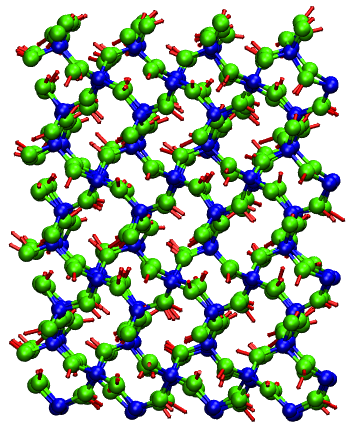}
\caption{\label{fig:dipolesquartz} Snapshot of a representative configuration obtained for BeF$_2$ in the $\alpha$-quartz structure. Blue: Be$^{2+}$ ions, Green: F$^-$ ions, Red: Vector showing the direction and sense of the induced dipoles on F$^-$ ions. Snapshot obtained using the VMD program.~\cite{vmd}}
\end{center}
\end{figure}

In order to illustrate this effect, we show in figure \ref{fig:dipolesquartz} the direction and sense of the induced dipoles on the F$^-$ ions in the $\alpha$-quartz structure of BeF$_2$. It appears clearly that this dipole points towards the bisector of the Be-F-Be angle. In other words, the electronic cloud of the anion is shifted with respect to the position of the nuclei, which effectively screens the cation-cation Coulombic repulsion.
In the framework of the RIM this screening could be mimicked through the use of partial charges; this approach will be discussed in the next subsection, but we immediately see that it suffers from transferability issues.
When studying a family of ionic compounds such as molten fluorides of varying compositions, the fluoride-fluoride interaction parameters should always be kept the same. This requirement holds for the Coulombic interaction parameters, and this can only be achieved by using formal charges. Such a transferable rigid ion model (TRIM) was proposed by Woodcock {\it et al.} for a series of systems including KCl and ionic liquids of MX$_2$ stoichiometry which have glass-forming ability (BeF$_2$, ZnCl$_2$, and SiO$_2$).~\cite{woodcock1976a} This model has proven very useful for understanding important physical aspects of the pure materials,~\cite{agarwal2007a,agarwal2009a,jabes2010a} but the transferability of rigid ion models across a wide range of mixture compositions has never been proven.

In recent years, a good deal of effort has been devoted to the study of molten fluorides because of their potential use as a solvent in the molten salt fast reactor.~\cite{delpech2009a} Experiments on molten fluorides are difficult because of the high melting points and their corrosive nature. A set of PIM interaction potentials has therefore been parameterized on
the basis of first-principles electronic structure calculations.~\cite{heaton2006a} These potentials are therefore {\em predictive} in the sense that no empirical information is used in their construction. The parameterization is based on a generalized ``force-matching" method. A suitable condensed-phase ionic configuration is taken
from a molecular dynamics simulation using some approximate force-field for the
material of interest. Typically a hundred ions would be used in periodic boundary
conditions. The configuration is then input to a planewave density functional theory (DFT) electronic structure
program and an energy minimization carried out to find the ground-state electronic
structure. From the results of this calculation the force and dipole moment on each
ion is obtained, the latter by making use of the transformation of the Kohn-Sham
orbitals to a Maximally Localized Wannier Function (MLWF) set.~\cite{marzari1997a} The parameters in
the polarizable potential are then optimized by matching the dipoles and forces
from the potential on the same ionic configuration to the \textit{ab initio} values.~\cite{heaton2006a} If necessary the process may be iterated, by using the fitted potential to generate a new ionic configuration to input to the \textit{ab initio}
calculation. In this approach, the dispersion terms have to be determined separately because of the use of functionals (e.g. PBE~\cite{perdew1996a}) for the DFT calculations in which these effects are not properly taken into account -- Note that this difficulty will probably be overcome in the nearest future thanks to the development of improved functionals~\cite{vydrov2010b}. The resulting potentials may be used in much larger scale  molecular dynamics simulations to obtain the physical properties of interest.

\begin{figure}
\begin{center}
\includegraphics[width=9cm]{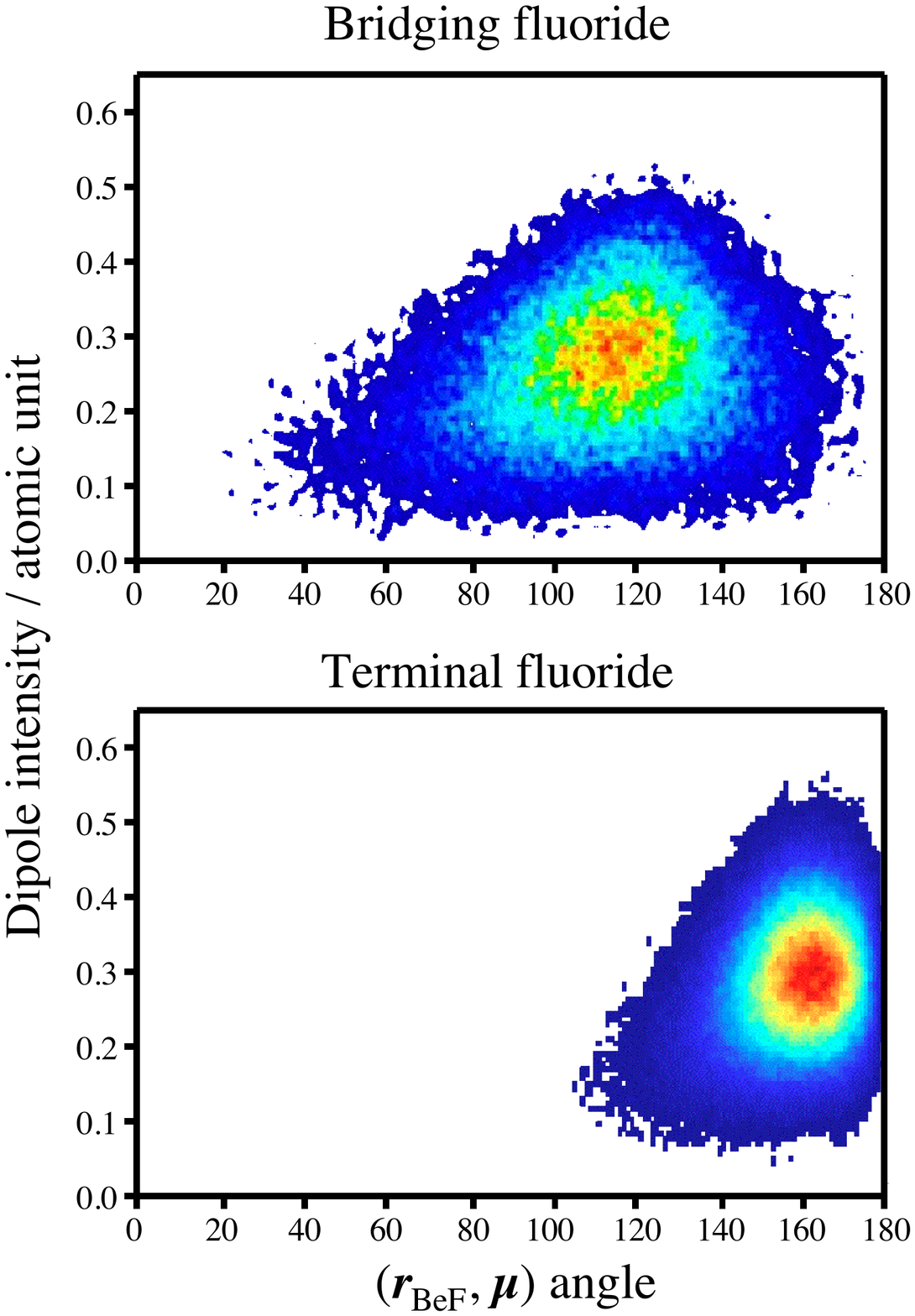}
\caption{\label{fig:dipolestatistics} Probability  to find an induced dipole on a fluoride anion with respect to its intensity and to the angle that it forms with the corresponding Be-F bond. Surfaces appearing in a red color correspond to the highest probability whereas the uncoloured ones correspond to a probability of zero. Data obtained for molten Li$_2$BeF$_4$ at a temperature of 873~K.}
\end{center}
\end{figure}

The procedure was validated on the LiF-BeF$_2$ mixtures. This choice was guided by the existence of a comprehensive experimental database for these materials as a legacy of the Molten Salt Reactor Program conducted in the US in the 60's. The predictive power of the simulations was tested against several experimental datasets without any empirical adjustments of the potentials. This included thermodynamic and transport data; at the atomistic structure level, the X-ray diffraction patterns~\cite{heaton2006a} as well the infrared and Raman spectra~\cite{heaton2006a,heaton2008a} were reproduced extremely well.  The picture of a network of BeF$_4$ tetrahedral entities that are connected by their corner has emerged. Depending on the BeF$_2$ concentration, the proportions of the various fluoroberyllate anions (BeF$_4^{2-}$, Be$_2$F$_7^{3-}$, Be$_3$F$_{10}^{4-}$ {\it etc}) could be quantified.~\cite{salanne2006a,salanne2007b} The important role of the induced dipoles in the transferability of the LiF-BeF$_2$ potential is highlighted in figure \ref{fig:dipolestatistics}. It shows the probability  to find an induced dipole on a fluoride anion with respect to its intensity and to the angle that it forms with the corresponding Be-F bond. Surfaces appearing in a red color correspond to the highest probability whereas the white ones correspond to a probability of zero. Two sets of data have been separated: The top panel correspond to the bridging fluorides, which are shared by two Be$^{2+}$ atoms. This figure is in agreement with the picture provided on figure \ref{fig:dipolesquartz} (which is normal since all the fluoride anions are bridging ones in $\alpha$-quartz BeF$_2$): The most likely angle is of around 120 degrees, i.e. when the induced dipole is directed along the Be-F-Be angle bisector. The bottom panel corresponds to the terminal fluoride, which are linked to one Be$^{2+}$ only. In that case the induced dipole is almost always directed along the Be-F bond, with a most likely angle of 160 degrees. The induced dipoles therefore allow the fluoride anions to adapt to distinct structural environments, which is not possible when using simpler pair-potentials only.

The validation of the first-principles procedures on the LiF-BeF$_2$ mixtures has allowed us to adopt a predictive strategy for physico-chemical properties that hitherto remain unknown despite their importance in establishing industrial proceses involving molten salts. This was done recently for ZrF$_4$ mixtures, for which a potential that was developed to predict thermodynamic properties~\cite{salanne2009a,salanne2009b} was successfully used to interpret high-temperature EXAFS spectroscopy data.~\cite{pauvert2010a,pauvert2011a} The structure of molten lanthanide chlorides, which involve trivalent cations, has also been elucidated thanks to the use of PIM potentials. Those were able to reproduce the neutron diffraction patterns~\cite{hutchinson1999a,hutchinson1999b,hutchinson2001a} as well as Raman~\cite{madden2004a,glover2004a} and EXAFS~\cite{okamoto2010a} spectrocopy data. In conclusion, it appears that taking the polarization effect into account is mandatory in any structural study of ionic materials involving multivalent species.

\subsection{On the use of partial charges in effective potentials}

As outlined above, in MX$_n$ compounds the main role played by the polarization of the anion X$^-$ on the local structure is to screen the strong repulsive electrostatic term between two cation M$^{n+}$ in the coordination shell of the anion, which affects the M-X-M angle and may influence the connectivity of the cation-centred coordination complexes (corner- or edge-sharing). Of course the drawback is an increase in the computational cost (mainly due to the self-consistent determination of the induced dipoles at each time step), so that it is tempting to find an "effective" way to account for it in the framework of the RIM. The simplest way to perform this is to reduce directly the cation-cation Coulombic interaction by using partial charges for the ions. This is routinely done in the case of molten silicates~\cite{guillot2007a,guillot2007b} and other oxide-based compounds. Most of the time, the parameters of the potential (including partial charges) are fitted to reproduce bond length and angles,~\cite{vanbeest1990a,oeffner1998a,carre2008a} so that these potentials can provide a good representation of the structure. We will show in the next section that it is much more difficult to simultaneously reproduce the dynamical properties. The polarizabilities of the ions gives rise to a value for the high (optical) frequency dielectric constant ($\epsilon_{\infty}$) greater than unity and at large interionic separations. The effect is to reduce the strength of the effective coulombic interactions by a factor of $\epsilon_{\infty}^{-1}$ \cite{grayweale2003a}. It is noteworthy that the reduction in the charges required in the successful silicate pair potentials is much larger than would be anticipated from this effect alone (see below).

\begin{figure}
\begin{center}
\includegraphics[width=9cm]{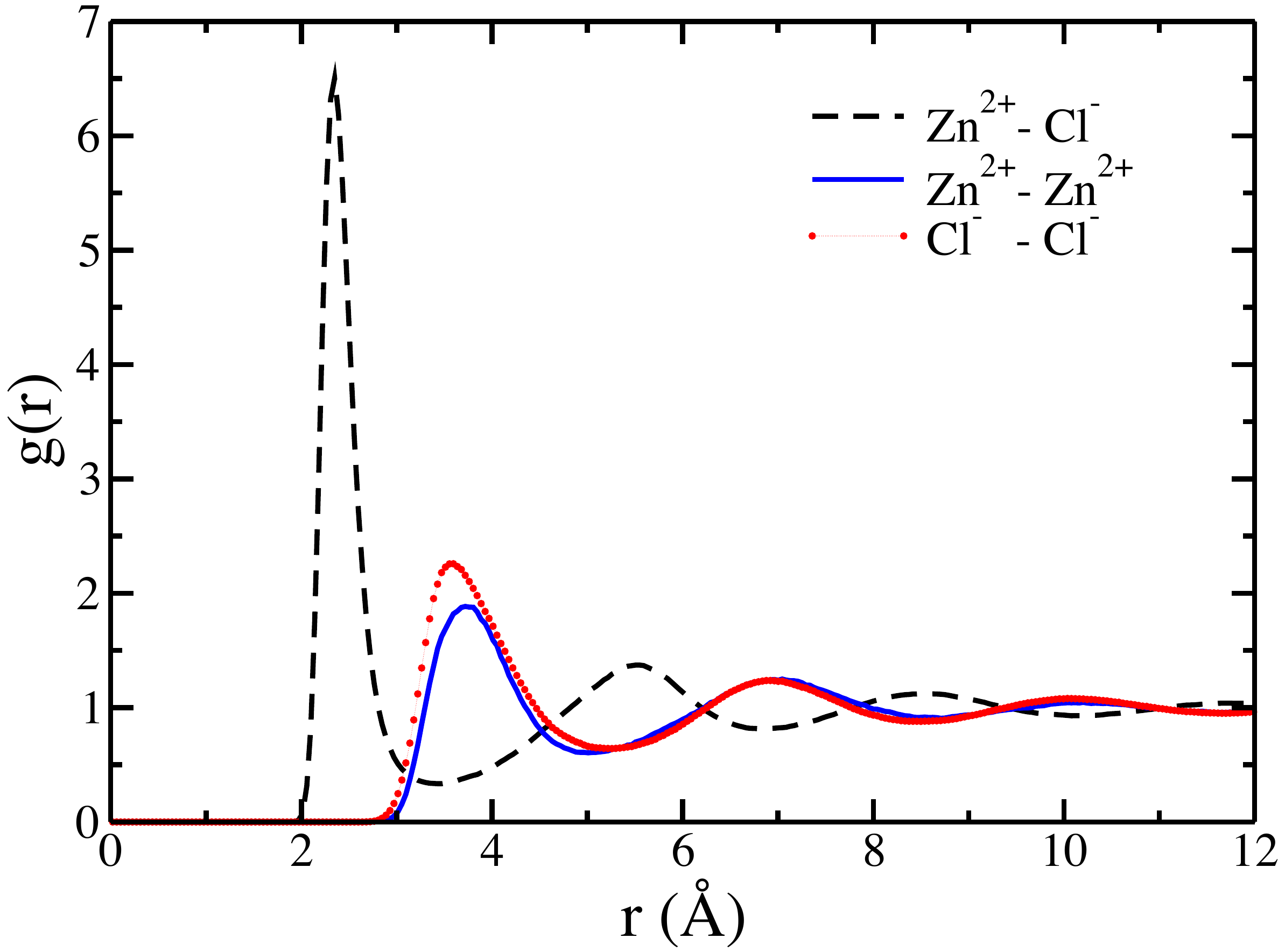}
\caption{\label{fig:rdfzncl2} Partial radial distribution functions in pure ZnCl$_2$ at 1000~K.}
\end{center}
\end{figure}

Much less literature is available in the case of halide compounds. One of the best documented examples for which the polarization effects are important is molten ZnCl$_2$. This system became the focus of attention when the first X-ray~\cite{triolo1981a} and neutron~\cite{biggin1981b} diffraction works appeared. Both studies showed the existence of intermediate range order in the liquid as manifested by the presence of a first sharp diffraction peak (FSDP) at a scattering vector $k$~=~1~\AA$^{-1}$ in the diffraction pattern. The first interpretation of their data by Biggin and Enderby led them to conclude that this FSDP was due to the Zn-Zn correlations.~\cite{biggin1981b} Although this conclusion has been challenged,~\cite{neuefeind2001a} it was confirmed recently by Zeidler {\it et al.} who have remeasured carefully the full set of partial structure factors.~\cite{zeidler2010a} The main particularity of their extracted RDFs is the position of the first peak for the Zn-Zn partial function, which occurs at the same position as the Cl-Cl one despite the larger charge on the cation.

Here we will only concentrate on the molecular dynamics simulation work which has been devoted to this compound. As already mentioned,  the TRIM potential developed by Woodcock {\it et al.} was able to provide the correct Zn-Cl first-neighbour distances, together with a structure organized in a tridimensional network. When the neutron data became available, it was clear that the Zn-Zn distance was too long because of the lack of screening of the Coulombic interactions. Gardner and Heyes tried to use several sets of partial charges without improving the situation.~\cite{gardner1985a} A RIM with an unphysically large dispersion attraction between the Zn cations does bring about the shift in the first peak of the Zn-Zn RDF, but at temperatures and densities where real ZnCl$_2$ is highly fluid, the simulated "fluid" is glassy.  Including the polarization effects in a PIM with formal charges gives the correct RDFs for the melt, which are shown on figure \ref{fig:rdfzncl2}, reproduces the intermediate-range order seen in the first sharp diffraction peak ~\cite{sharma2008a}, and also gives good dynamical properties. The agreement with experiment extends to mixtures with alkali chlorides~\cite{wilson1994a}.

\subsection{Polarization effects on the dynamics}

In condensed phases, the dynamical processes with the fastest relaxation time are associated with the vibrations of the atoms. In crystals, vibrational properties may be studied from the phonon dispersion relationship while in amorphous compounds (glasses, liquids) infrared and Raman spectroscopies provide information on the characteristic vibrational motions. As we have already indicated, in many melts, especially those with polyvalent cations, relatively long-lived, quasi-molecular coordination complexes form and these give rise to well-resolved characteristic vibrational bands.\cite{papatheodorou1977a,photiadis1998a,dracopoulos2001a,glover2004a} On longer time scales, in the liquid state or in superionic solids,~\cite{marrocchelli2009a,norberg2011a,marrocchelli2011a} the dynamics is characterized by the diffusion of ionic species in response to local chemical potential gradients. These transport properties are then quantified by a diffusion coefficient for each species present in the system.~\cite{hansen-livre} The diffusion coefficients, conductivity and viscosity in the melt are strongly influenced by the lifetimes of the coordination complexes and by the degree to which these complexes are linked together to form a network.~\cite{brookes2005a,salanne2007b}
 The inclusion of the polarization effects in the interaction potential is crucial for describing these complexes and their network-forming tendencies correctly.

All the vibrational and diffusive properties can be rather straightforwardly extracted from a molecular dynamics simulation. Even if the use of partial charges in rigid ion models may suffice to bypass the inclusion of an explicit polarization term in the study of structural properties of ionic compounds, to also reproduce the dynamic properties with such a structurally optimized model is far more demanding. In order to illustrate this, we take the example of germania, GeO$_2$, which is a close structural analog of silica.~\cite{micoulaut2006b,salmon2007a} For this system, Oeffner and Elliott have obtained a first set of parameters for a rigid ion model by using a two step procedure.~\cite{oeffner1998a} In a first stage, they calculated the {\it ab initio} potential energy surface of a Ge(OH)$_4$ cluster, on which they could fit one hundred possible set of potentials. Among those, they chose the potential that better reproduced the characteristic bond lengths and bond angles for the $\alpha$-quartz GeO$_2$ structure. This potential involved for example partial charges of +1.5~$e$ and -0.75~$e$ for the Ge and O atoms respectively. It was used by Hawlitzky {\it et al.} \cite{hawlitzky2008a} in simulations of the liquid state, and was shown to provide diffusion coefficients compatible with the experimental data (which consists in a series of viscosity measurements, from which the diffusion coefficients were estimated using Eyring equation). Nevertheless, as soon as vibrational properties are concerned, important discrepancies have been observed. For example, the vibrational density of states predicted too high frequencies for all the bands found for $\alpha$-quartz GeO$_2$, which led  Oeffner and Elliott to propose a second set of parameters, in which all the interactions were rescaled (for example, the new partial charge is of 0.94174~$e$ for Ge atom).~\cite{oeffner1998a} This potential now provided a better description of the vibrational properties, but then its use in liquid state simulations yielded values of the diffusion coefficients which were one order of magnitude too high.~\cite{micoulaut2006a,marrocchelli2009a}

Interaction potential parameters were obtained for a PIM (with formal valence charges) of GeO$_2$ by Marrocchelli {\it et al.} by force-fitting to first-principles DFT calculations~\cite{marrocchelli2009a}. The analytic expression for the repulsion term used in that study differed slightly from the one given in equation \ref{eq:rim}  in order to enhance the stability of the potential against numerical problems in high temperature simulations. This potential was shown to provide diffusion coefficients in agreement with the original Oeffner-Elliott potential together with very good vibrational properties. The latter could be tested by calculating the infrared spectrum in the glassy state. In fact, the inclusion of polarization effects for the oxide ions in the model may influence the predicted spectrum in two ways.~\cite{wilson1996a} First, the interactions of the oxide ion dipoles may alter the
local structure of the network and the strength of the bonds, which may introduce a
shift of the vibrational frequencies. Second, the induced dipoles will themselves
be responsible for absorption, as they too contribute to the total polarization
fluctuations. The absorption coefficient in the presence of these extra moments is calculated
from the imaginary part of the total dielectric function $[n(\nu)\alpha(\nu)=
2\pi \nu \Im(\epsilon(\nu))]$, which can be determined following Caillol {\it et al.} \cite{caillol1989a,caillol1989b} as
\begin{eqnarray}
\epsilon(\nu)-\epsilon_\infty&=&\frac{\beta}{3 \epsilon_0 V}\left(\langle {\bf M}(0)^2\rangle +2\pi \imath \nu \langle {\bf M}\cdot{\bf M}\rangle_\nu  \right.  \\
&&\left.+2\langle {\bf M}\cdot{\bf J}\rangle_\nu+\frac{\imath}{2\pi\nu}\langle {\bf J}\cdot{\bf J}\rangle_\nu \right) \nonumber
\label{ir}
\end{eqnarray}

where
\begin{equation}
\langle {\bf J}\cdot{\bf J}\rangle_\nu = \int_0^\infty e^{2\pi\imath\nu  t} \langle {\bf J}(t)\cdot {\bf J}(0)\rangle {\rm d}t,
\end{equation}
${\bf J}(t)$ is the charge current ${\bf J}(t)=\sum_{i} q^i{\bf v}^i(t)$ and ${\bf M}(t)$ is the total system induced dipole moment, ${\bf M}(t)=\sum_{i}{\boldsymbol \mu}^i(t)$. The agreement was found to be very satisfactory, with a good correspondence of peak frequencies with experiment and a much better distribution of intensity across the spectrum than obtained by omitting the induced dipole terms in equation \ref{ir}.

\begin{figure}[htbp]
\begin{center}
\includegraphics[width=9cm]{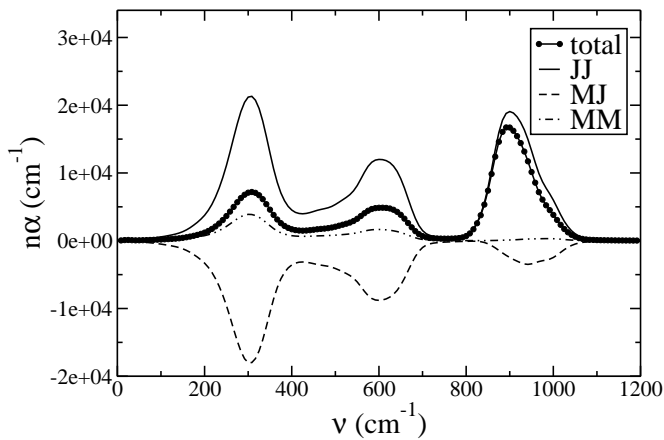}
\end{center}
\caption{Decomposition of the total infrared absorption spectrum of glassy GeO$_2$ in its three component parts (JJ: charge-charge, MJ: charge-dipole, MM: dipole-dipole). }
\label{fig:decompo}
\end{figure}

The importance of polarization effects in determining the relative intensities of
the bands can be demonstrated by separating the various contributions to the
absorption spectrum. In the case of glassy silica \cite{wilson1996a} and beryllium
fluoride \cite{heaton2006a}, it was observed that the interference between the
induced dipoles and permanent charge contributions to the total polarization,
contained in the $\langle {\bf M}\cdot{\bf J}\rangle_\nu$ cross term, is
responsible for the changes of the relative intensities. Figure \ref{fig:decompo} shows
the decomposition of the absorption spectrum in the case of glassy GeO$_2$. The two main
contributions are the charge fluctuation and the cross term; both show  bands at
the same intensities. Note that the cross-term strongly reduces the intensity of
the two low frequency bands relative to that which would be obtained from the
charge fluctuations alone, whereas for the high-frequency band the cancellation is
much weaker. For the high frequency band the relationship of the charge-charge and
cross-terms is different to that found previously for SiO$_2$ and
BeF$_2$\cite{wilson1996a,heaton2006a}. In the latter cases, the charge-charge and
cross-terms had the same sign, so that the net band intensity was slightly larger
than that predicted by the charge-charge term alone. This difference in the
behaviour of the calculated spectra might arise from the inclusion of an
anion-anion damping term in the polarization part of the interaction potential
here, which was not the case for the previous studies.

The {\it ab initio}-parameterized interaction  potential was subsequently used to study the behavior of GeO$_2$ under pressure in the glassy and liquid states.~\cite{marrocchelli2010a} It could predict a smooth transition from a tetrahedral to a octahedral network with a significant number of pentacoordinated germanium ions appearing over an extended pressure range, in agreement with the most recent experimental data.~\cite{vaccari2009a,drewitt2010a,mei2010a}

Polarization effects also influence the long-time dynamics. Even in alkali halides, where the structural consequences of including polarization effects are not large, diffusion coefficients tend to be increased. This has been shown, for example, in studies of mixtures of LiCl and KCl~\cite{morgan2004a} of various compositions which were simulated by using both a rigid and a polarizable ion model in which all the repulsion, dispersion and ionic charges were kept exactly the same and, in the PIM case,  a realistic polarizability of the Cl$^-$ ion was used. The comparison of the two sets of diffusion coefficients showed that the effect of including polarizability is to make the system more mobile. Interestingly, the largest increase concerns the Li$^+$ ions, although these ions are not polarizable. The calculated diffusivities for the pure melts appeared to be much closer to the experimental values for the PIM. Similar calculations were performed for a room temperature ionic liquid, namely the 1-ethyl-3-methylimidazolium nitrate.~\cite{yan2010a,yan2010b} Here again, the effect of including polarization effects was to enhance the fluidity of the liquid. This reflected on the whole set of transport coefficients: With the polarizable model the diffusion coefficients and ionic conductivities were increased, and consistently the viscosity was decreased. Far more dramatic increases in fluidity are seen in systems with polyvalent cations, here the issue is complicated by making a direct comparison between a PIM and an appropriate RIM. In many cases, simply omitting the polarization terms from a PIM gives an RIM which has hopeless structural and thermodynamic properties (which is not the case for alkali halides). It seems better to compare a PIM with an independent RIM which has been optimized, so far as possible, to reproduce the structure. In simulations of UCl$_3$, \cite{okamoto2005b} for example, such a comparison showed order of magnitude increases in the diffusivities of both ions with the PIM reproducing the measured conductivity and viscosity very well.

From the mechanistic point of view, these results are well understood in terms of the screening of the charge-charge interactions by the induced dipoles. The polarization energy always lowers the total coulombic energy - it is calculated by minimizing the latter (equation \ref{minimization}). The effect is always largest for ionic configurations in which the ions are most strongly polarized, which is the case when they sit in an asymmetric coordination environment. If one visualizes diffusion as ions hopping between locally metastable configurations, it is easy to imagine that the transition states in such hops are more asymmetrically coordinated and hence that polarization will lower the energy more at the barrier tops than at the minima.

The picture of polarization effects lowering a barrier to structural relaxation accounts for the influence on those transport coefficients, like diffusion or viscosity, which depend on such events. The situation is somewhat more complicated for the thermal conductivity which measures a material's ability to conduct heat, and is given, for a binary mixture, by~\cite{sindzingre1990a}
\begin{equation}
\lambda = T^{-2}\left(L_{EE}-\frac{L_{EZ}^2}{L_{ZZ}}\right).
\label{eq:lambda1}
\end{equation}
\noindent where
\begin{equation}
L_{ab} = \frac{1}{3Vk_B}\int_0^\infty\langle{\bf j}_a(t)\cdot{\bf j}_b(0)\rangle {\rm d}t.
\end{equation}
\noindent For systems consisting of more than three charged species, different expressions need to be derived for $\lambda$.~\cite{salanne2011a} Here the quantities involved are the charge current ($a~=~Z$), which is defined as previously, and the energy current ($a~=~E$). The computation of the latter quantity requires to take special care due to the use of the Ewald summation technique. The first expression was derived by Bernu and Vieillefosse in their study of the transport coefficients of the one-component plasma,~\cite{bernu1978a} and we extended this work to the case of potentials including polarization effects.~\cite{ohtori2009a}

\begin{table}[t]
\begin{center}
\begin{tabular}{c c c c c} \hline
System & T (K) & $\lambda^{\rm RIM}$  & $\lambda^{\rm PIM}$  & $\lambda^{\rm exp}$ \\
\hline
LiCl & 1200 & 0.841 & 0.643 & 0.534\\
NaCl & 1300 & 0.581 & 0.509 & 0.478\\
KCl  & 1300 & 0.387 & 0.343 & 0.345\\
\hline
\end{tabular}
\end{center}
\caption{\label{tab:thermalconductivity} Values of the thermal conductivity (in Wm$^{-1}$K$^{-1}$) for a series of
molten chlorides obtained using PIM and RIM interaction potentials, and from experiments.~\cite{nagasaka1992a}}
\end{table}

The values obtained for a series of molten chlorides (LiCl, NaCl, KCl) in the latter study~\cite{ohtori2009a} are summarized in table \ref{tab:thermalconductivity}. In this work, the PIM parameters were obtained wholly from first-principles DFT calculations, and the RIM simply corresponds to the same potential where the polarization effects are omitted. We observe that the PIM systematically predicts lower thermal conductivities, which are in much better agreement with the experimental data.~\cite{nagasaka1992a} The non-polarizable model yields similar values to the famous Tosi-Fumi potentials,~\cite{tosi1964a,fumi1964a,galamba2007a} which are known to give the correct structural properties of the MX molten salts in general. This shows again that the inclusion of polarizability is crucial for a good representation of the thermal relaxation. The thermal conductivity is affected by different aspects of the ionic dynamics than the other transport coefficients, such as the viscosity or ionic conductivity; this quantity can therefore be used to test the interaction model in different ways. For the set of alkali halides, the first-principles-determined PIM potentials predicted values for \emph{all} transport properties with the largest discrepancy being about 10\% of the measured value.~\cite{ohtori2009a}

\subsection{Polarization effects on the interfacial properties}

\subsubsection{Liquid-vapor interface}

The simplest interface involving an ionic liquid that one may consider is the one with its own vapor. The experimental vapor pressure of these systems is very low, which means that from the point of view of the molecular simulation this interface is in reality an ionic liquid -- vacuum one. The simulation cell now has dimension $L\times L\times L_z$, where $L_z=D+L_{\rm vacuum}$. Due to the use of vacuum periodic boundary conditions in the Ewald summation instead of conducting boundary conditions for the calculation of long-range electrostatic interactions, additional terms have to be added to the energy, forces and pressure tensor.~\cite{yeh1999a} It is also necessary to perform an Ewald summation of dispersion interactions,~\cite{karasawa1989a}  to avoid substantial truncation effects.~\cite{aguado2001a}

\begin{figure}[htbp]
\begin{center}
\includegraphics[width=9cm]{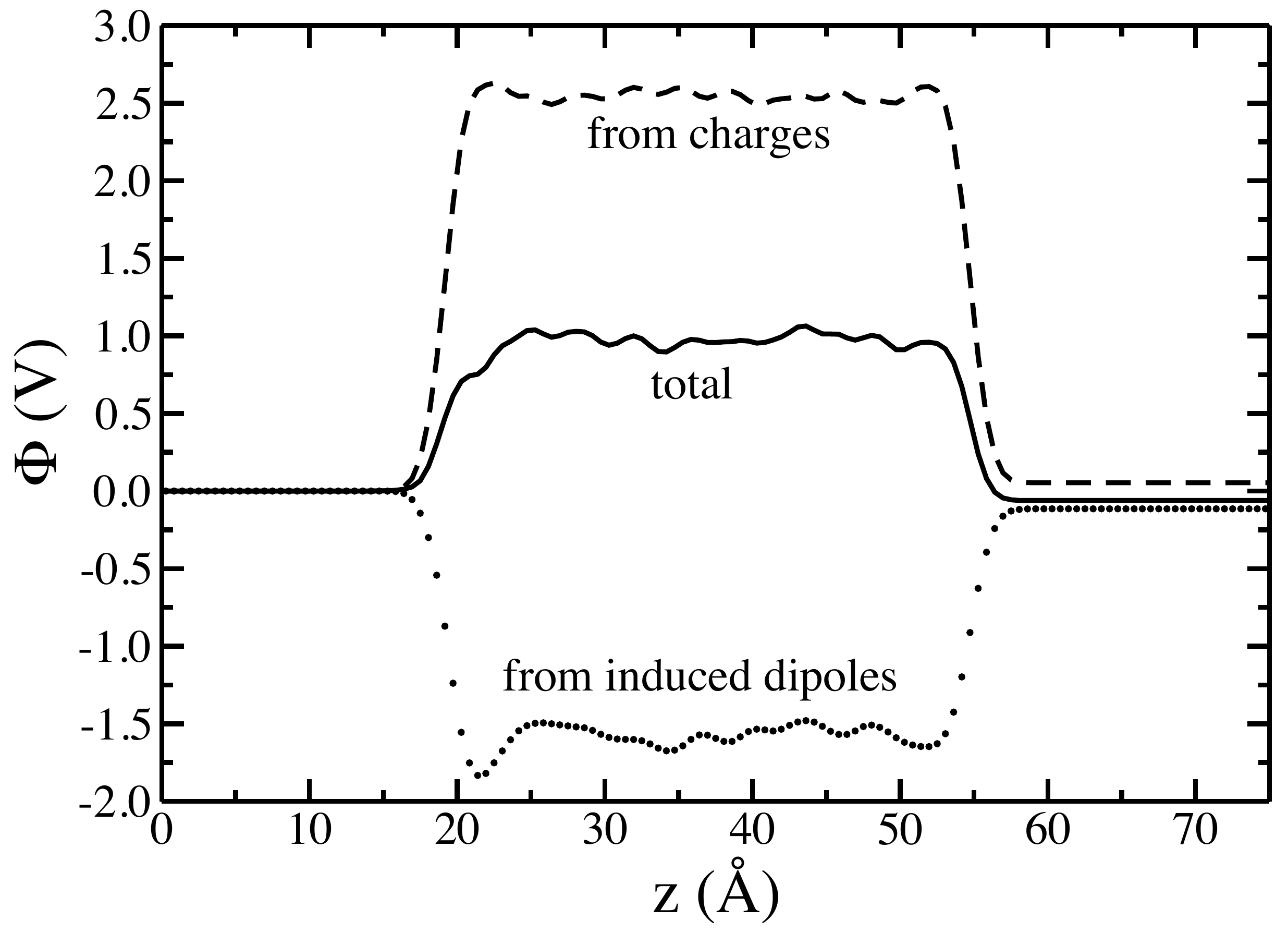}
\end{center}
\caption{Electrostatic potential profile across the liquid-vapor interface in molten LiBeF$_3$ at 1060~K. The contributions due to the charges and to the induced dipoles are also shown.}
\label{fig:electrostaticpot}
\end{figure}

The electrostatic potential difference across the interface contains two contributions,
\begin{eqnarray}
\Delta \Phi(z)&=&\Delta \Phi_q(z)+\Delta \Phi_\mu(z)\\
&=&\frac{1}{\epsilon_0}\left(-\int_{z_0}^z {\rm d}z'\int_{z_0}^{z'} \rho_q(z''){\rm d}z''+ \int_{z_0}^z \rho_\mu(z'){\rm d}z'\right)
\end{eqnarray}
\noindent which respectively correspond to the distribution of charges and induced dipoles. In this expression $z_0$ corresponds to a point in the vapor region. The importance of the two terms is showed on figure \ref{fig:electrostaticpot}, which corresponds to molten LiBeF$_3$ at a temperature of 1060~K.~\cite{salanne2007a} In this system, we observed the stabilization of the fluoroberyllate species at the interface, leading to a local enhancement of the Be$^{2+}$ ions concentration. This segregation leads to important charge-separation effects at the interface, which explains the very large value of the charge term. An electronic dipole moment is also created on the fluoride atoms at the interface, which results in a contribution opposite in sign, thus softening the electrostatic potential variation. The total electrostatic potential difference between the vacuum and the ionic liquid is of approximately 0.98~V.

Although it appears clearly that without the inclusion of polarization effects the structure of the liquid -- vacuum interface of LiBeF$_3$ would differ a lot, no explicit comparison has been made for this system. Such a comparison of the RIM and PIM interaction potentials was performed in the case of KI, in a series of papers dedicated to a better understanding of the interfacial properties of molten salts.~\cite{aguado2001a,aguado2001b,aguado2002a} It was shown that polarization tends to increase the interfacial width by reducing slightly the bulk density and enhancing the ion density in the interfacial region. Stronger effects were observed for the surface tension, which can be obtained through the mechanical definition
\begin{equation}
\gamma = \frac{L_z}{2}\left( \langle \Pi_{zz} \rangle -\frac{1}{2}\langle \Pi_{xx}+\Pi_{yy} \rangle \right)
\end{equation}
\noindent where the various $\Pi_{\alpha\alpha}$ are the diagonal pressure tensor components. In the case of KI,  the effect of polarization is to reduce the surface tension by about 20~\%. By calculating the Coulombic, repulsion, dispersion and polarization components for the surface tension, Aguado {\it et al.} could show that the polarization component is small and positive, so that the effect of polarization on $\gamma$ is indirect: It is a consequence of the increase of the interfacial width.~\cite{aguado2001a} It is worth mentioning that the importance of polarization effects in the behavior of ions at the air -- water interface has also been the object of numerous studies.~\cite{jungwirth2002a,jungwirth2006a,chang2006a}

\subsubsection{The ionic liquid -- electrified metal interface}
The study of the interface which is formed when an ionic liquid is in contact with a metal is of particular importance; the interest derives from several sources. In the case of high temperature molten salts, understanding the corrosion mechanisms is essential for the design of Generation IV nuclear reactors~\cite{vergnes2002a,forsberg2003a,mathieu2006a,delpech2009a} as well as for the improvement of pyrometallurgical applications, such as the Hall-H\'eroult process - which is the major industrial process for the production of aluminium.~\cite{cassayre2007a,cassayre2010a} As for room-temperature ionic liquids, they are used as electrolytes in electrochemical supercapacitors,~\cite{simon2008a,largeot2008a,lin2009a} batteries, and fuel cells.~\cite{armand2009a} In all these applications, a better understanding of the properties of the ionic liquid in the vicinity of the charged surface would be beneficial.

The study of such interfaces by molecular dynamics simulations has been made possible with the introduction of a method to represent the polarization of a model metallic electrode which is maintained at a controlled electric potential difference.~\cite{reed2007a} The system is periodically replicated in the plane parallel to the electrodes. The metallic, constant potential condition is attained by minimizing a suitable energy function with respect to variable charges on the electrode atoms, following a procedure suggested by Siepmann and Sprik in a different context.~\cite{siepmann1995a}

A suitable interaction potential was obtained, on the basis of small scale DFT calculations, for a system consisting of a molten LiCl electrolyte and a solid aluminium electrode. In addition to the four terms already present in the PIM, it was found necessary to include an additional short-range term between the metal atoms and the ions,
\begin{equation}
V^{SR}=\sum_{i=1}^N\sum_{j=1}^M A_{SR}^{ij}{\rm e}^{-a_{SR}^{ij}(r^{ij}-d)^2}\hat{{\bf r}}^{ij} \cdot {\boldsymbol{\mu}}^i,
\label{metal-ion}
\end{equation}
\noindent where $\hat{{\bf r}}^{ij}$ is a unit vector, in order to obtain a satisfactory representation of DFT-calculated dipoles and forces ($N$ and $M$ respectively are the number of  ions in the liquid and of atoms in the metal) for the ions close to the electrode surface.~\cite{pounds2009a} This term is directed along the interatomic separation; it behaves like an electric field which distorts the charge cloud of a melt ion due to overlap-mediated interactions with the electrode atoms. The functional form in equation \ref{metal-ion} is somewhat arbitrary, as the range of atom-ion separations at which these interactions are sampled, in the immediate vicinity of the electrode surface, is small and insufficient to determine the shape of the full function. This form allows the forces and dipoles to be adequately represented in the physically significant regions without bad computational problems in regions which are physically irrelevant because of the repulsive interactions between electrode and ions.

\begin{figure}[htbp]
\begin{center}
\includegraphics[width=9cm]{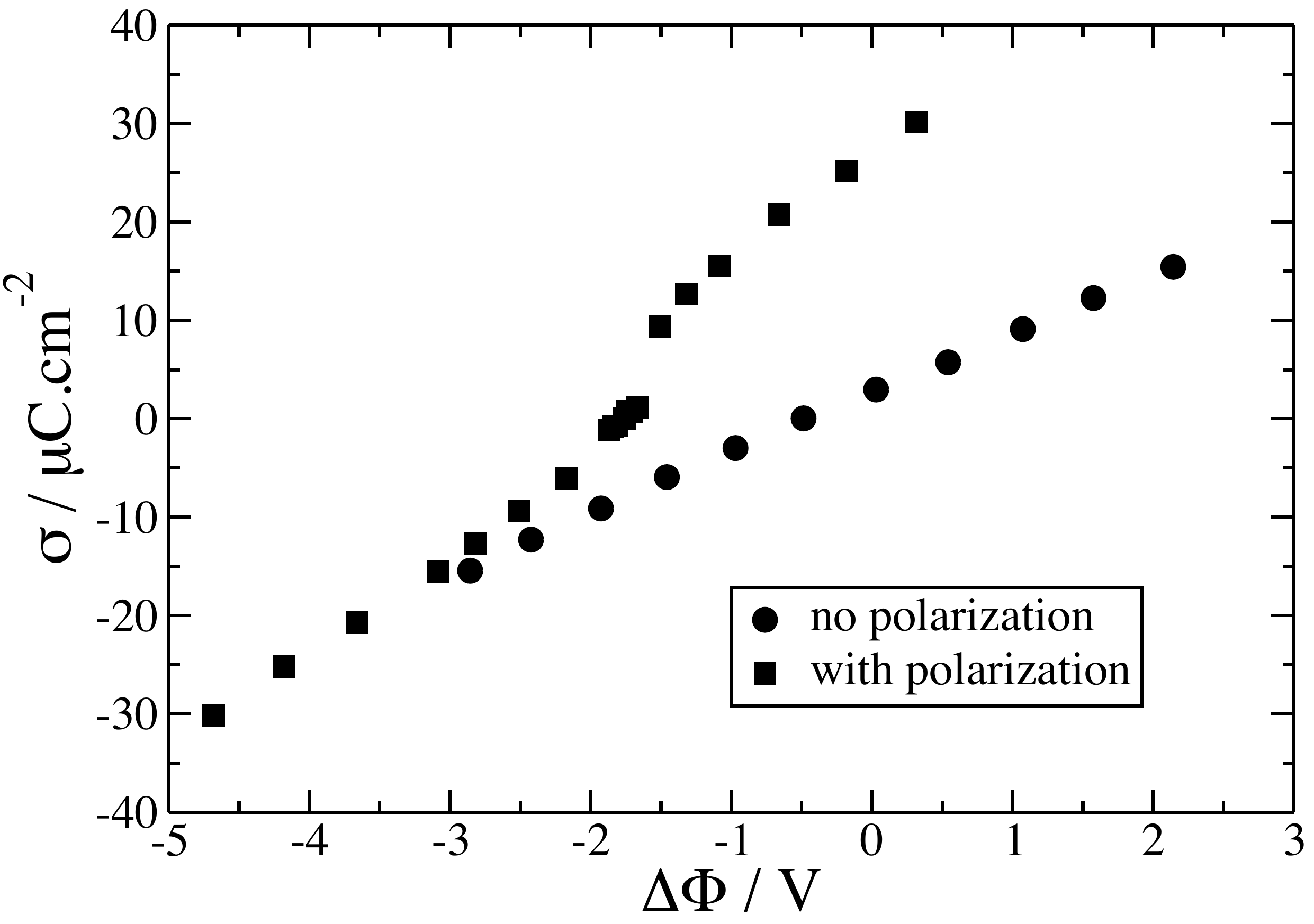}
\end{center}
\caption{Variation of the metal surface charge with the potential drop across the electrified aluminium -- molten LiCl interface.}
\label{fig:capacitance}
\end{figure}

The model was then used in simulations of this system for various values of the electrical potential difference applied between the electrodes. The first relevant quantity that can be extracted from such simulations is the electrostatic potential profile. It differs from the one shown in figure \ref{fig:electrostaticpot} because the potential in the electrodes is not 0~V anymore, unlike for the liquid-vapor interface. It is also possible to calculate the average charge of the electrodes $\sigma$. Then the potential of zero charge (PZC) corresponds to the potential drop across the electrode -- liquid interface for which the accumulated charge on the electrode surface is zero on average.
\begin{equation}
\Delta \Phi_{\rm PZC}=\Delta \Phi (\sigma = 0)
\end{equation}
\noindent The calculated PZC cannot be compared to experimentally measured quantities since the latter are always measured with respect to a reference electrode whereas we are dealing with absolute potentials. The variation of the surface charge with respect to $\Delta \Phi$ was calculated in the case of the electrified aluminium -- molten LiCl interface with and without the inclusion of polarization effects for the chloride anions (the lithium cations are not polarizable).~\cite{tazi2010a} Three important differences are observed when polarization effects are included: First, the PZC goes toward a more negative value. Second, the differential capacitance, which is given by the derivative of this function,
\begin{equation}
C=\frac{\partial \sigma}{\partial\Delta\Phi}
\end{equation}
\noindent becomes higher, switching from 6.1~$\mu$F~cm$^{-2}$ to 10.2~$\mu$F~cm$^{-2}$. The analysis of the induced dipoles distribution showed that they bring an additional screening of the electrode charge,~\cite{tazi2010a} and that the magnitude of the dipole increases with the electrode potential.

But the most important difference between the two sets of data is the presence of a discontinuity when the polarization effects are included. This increase is due to a potential-driven ordering transition in the adsorbed layer of electrolyte, which occurs at the PZC. The adsorbed layer switches from a fluid-like structure to a 2-dimensional crystalline one which is formed through an epitaxial mechanism to adapt to the electrode surface structure. This crystalline structure did not form when the RIM was used, which shows again that taking polarization effects into account is obligatory when studying the interfacial properties of molten salts - even for an alkali halide, where the polarization effects on the structure of the melt itself are very small. For RTILs, the situation is much more complex:~\cite{vatamanu2010a,vatamanu2011a} The internal charge distribution within the molecular ions already induces additional screening of the electrode charges and electronic polarization effects may play a secondary role. For example, a simple RIM was able to yield an ordered surface structure for the adsorption of 1-butyl-3-methylimidazolium  hexafluorophosphate on a graphite electrode.~\cite{kislenko2009a}

\section{Conclusions and perspectives}

In this topical review we have given an overview of the importance of polarization effects on the physico-chemical properties of ionic materials, with particular focus on their melts. Most of the information was extracted from molecular simulation studies, in which the choice of the model ({\it e.g.} the rigid ion model versus the polarizable ion model) will determine whether these effects are taken into account. In most halide and oxide materials, the most polarizable species is the anion. The creation of induced dipoles affects the thermodynamics of the system in different ways. From the structural point of view, although polarization does not seem to play an important role for simple alkali halides, it becomes very important as soon as multiply charged cations are involved. There, the induced dipoles screens the repulsive cation-cation Coulombic interaction, thus allowing the formation of bent M-X-M angles and thereby influencing the character of the networks which form through the linkage of cation-centred coordination complexes . Most of the attempts to introduce this screening effectively by using partial charges failed in predicting the correct structural properties. Concerning the dynamic properties, polarization seems to systematically increase the fluidity of the liquids, which results in higher diffusion coefficients for all the ions (including the non-polarizable ones such as Li$^+$) and the electrical conductivity as well as in a lower viscosity. Finally, the recent work devoted to the study of the interfacial properties of molten salts have shown the importance of the induced dipoles for the stabilization of particular surface structures such as the formation of an ordered layer of LiCl on the (100) surface of metallic aluminium.

Alongside the various qualitative points discussed within this manuscript we have also emphasized  the importance of reliable interaction potentials. A substantial effort was devoted to developing such potentials in recent years.~\cite{salanne2011c} The polarizable ion model, because it introduces additional degrees of freedom which mimic the response of the electronic structure of the ions to their changing coordination environments, can routinely be parameterized by fitting the predicted forces and dipoles to a large body of information generated from first-principles calculations. The resulting potentials are of first-principles accuracy, which means that molecular dynamics of ionic compounds can now be used  as a predictive tool under various conditions.

Yet it is important to notice that there is still room for improvement of the models. For example, some environmental effects occur on the electronic structure of the ions when the nature of the liquid is changed. In fluoride molten salts, for example, calculations of the condensed phase ionic polarizabilities have shown that the fluoride anion polarizability may shift a lot, passing from 7.8~a.u. in molten LiF to 11.8~a.u. in CsF.~\cite{salanne2008e}  This is due to differences in the confining potential, which affects the electron density around a given anion, and  originates from both Coulombic interactions and the exclusion of electrons from the region occupied by the electron density of the first-neighbor shell of cations.~\cite{fowler1984a,fowler1985a,jemmer1998a} When passing from one cation to another (for example in the series Li$^+$$\rightarrow$Na$^+$$\rightarrow$K$^+$$\rightarrow$Cs$^+$), two effects are then competing: On the one hand, the anion-cation distance increases, which results in a diminution of the confining potential, but on the other hand, the volume occupied by the cation electron density also increases, with an opposite effect on the confining potential. Here, the observed increase of polarizability with the size of the cation tends to show that the first effect is the most important. Indeed, the value obtained for CsF is approaching the free F$^-$ anion polarizability, which is 16~a.u. These effects have also been observed in similar calculations performed on solid oxides~\cite{heaton2006b} and protic solvents~\cite{buin2009a,salanne2011b}. Such a result means that in order to build a completely transferable interaction potential, the model should  allow the variation of the polarizability in response to the fluctuations of the environments. First steps have been made in that direction in the case of MgO only.~\cite{aguado2004a,aguado2005a}
\begin{figure}
\begin{center}
\includegraphics[width=9cm]{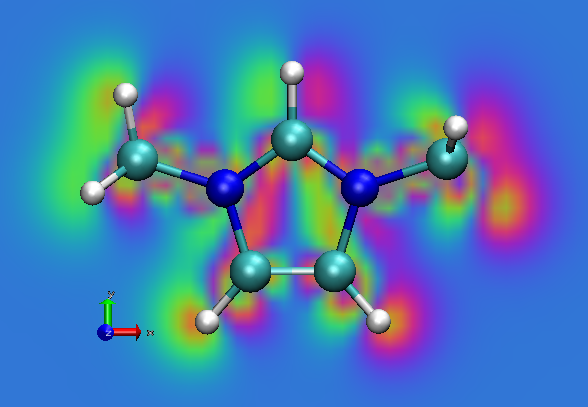}
\end{center}
\caption{Variation of the electronic density when an external field is applied to an isolated 1-ethyl-3-methylimidazolium cation. Green and red zones respectively correspond to an increase (decrease) of the electronic density with respect to the situation without the external field.}
\label{fig:rtil}
\end{figure}

In the future it is probable that the polarizable force fields will also systematically be used for room-temperature ionic liquids. Although several studies have recently been reported,~\cite{yan2004a,yan2006a,yan2010a,yan2010b,borodin2009a,bedrov2010a,salanneinpress} the general approach of most groups remains to discard polarization effects. In fact, the situation seems to be rather similar to the case of simple alkali halides, with a correct representation of the structural properties and a poor estimation of the transport properties in general.~\cite{yan2010a,yan2010b} From the technical point of view, the main difference between inorganic salts and room temperature ionic liquids arises from the molecular structure of the latter. As depicted on figure \ref{fig:rtil} where we plot the  variation of the electronic density when an external field is applied to an isolated 1-ethyl-3-methylimidazolium cation, there now are some intramolecular polarization effects that have to be taken into account. This can be done efficiently using the approach proposed by Thole,~\cite{thole1981a} which has now been introduced in numerous molecular dynamics simulation codes.

\bibliography{references}

\end{document}